\documentclass[12pt]{article}
\usepackage{amsmath}
\usepackage{graphicx}
\usepackage{enumerate}
\usepackage{natbib}
\usepackage{url} 
\usepackage{amsmath}
\usepackage{subfig}
\newcommand{\blind}{0}

\begin{document}

\def\spacingset#1{\renewcommand{\baselinestretch}%
{#1}\small\normalsize} \spacingset{1}


\if0\blind
{
  \title{\bf A cheap data assimilation approach for expensive numerical simulations}
  \author{Bijan Fallah\thanks{
    \textit{The author gratefully acknowledges German Federal Ministry of Education and Research (BMBF) as Research for Sustainability initiative (FONA); through Palmod project (FKZ: 01LP1511A). The computational resources were made available by the High Performance Computing Center (ZEDAT) at Freie Universit\"{a}t Berlin and the German Climate Computing Center (DKRZ). I acknowledge  GeoHydrodynamics and Environment Research (GHER) research group for making their OI code available for public. My special thanks go to Ingo Kirchner for his helpful comments on simulation set-up as well as the CCLM community.}}\hspace{.2cm}\\
    Institute of Meteorology, Freie Universit\"{a}t Berlin\\
    }
  \maketitle
} \fi

\if1\blind
{
  \bigskip
  \bigskip
  \bigskip
  \begin{center}
    {\LARGE\bf Title}
\end{center}
  \medskip
} \fi

\bigskip
\begin{abstract}
Using a very cheap Data Assimilation (DA) method, I show an alternative approach to classical DA for numerical climate models which produce a large amount of ``big data''. The problematic features of state-of-the-art high resolution Regional Climate Models are highlighted. One of the shortcomings is the sensitivity of such models to the slightly different initial and boundary conditions which could be corrected by assimilating scattered observational data. This method might help to reduce the bias of numerical models based on available observations within the model domain, especially for the time-averaged observations and the long-term simulations. 
\end{abstract}

\noindent%
\vfill

\newpage
\spacingset{1.45} 
\section{Introduction}
\label{sec:intro}
The concerns about the climate change require deeper learning about the uncertainties in future predictions of climate system. However, our cutting-age science appears helpless to describe processes which take place beyond the yearly cycle of the climate system. The main question in this regard will be \textit{why there is an urgent need for such studies?} It is now well known that many of long-term processes (millennial scale) in the climate system may have a large impact on predicting the climate even in shorter time scales (decadal, yearly scales) \citep{Evans2013,Acevedo2015,Fallah2015,Acevedo2016,Latif2016}. According to \cite{Lovejoy2016}, the atmosphere shows an extremely changing behaviour $-$ \textit{``more than 20 orders of magnitude in time and 10 in space: billions of years to milliseconds and tens of thousands of kilometers to millimeters''}. There exist several challenging open questions in the field of future climatic change: How variable is the future climate?, What is the probability of 
extreme future climate changes?, How would this impact the sea level rise?\\

\subsection{Challenges in long-term climate investigations}

Available observational climate data sets are the most accurate source of knowledge about the climate system. However they suffer from ill structural conditions: (i) time span of these data sets are usually less than a century, (ii) they change their accuracy in time (environmental changes in station location, usage of different measuring devices, urbanization), (iii) they change their density in time (more stations are available online in recent years), (iv) more numbers of stations over land than over oceans. 

Climate proxy archives (tree ring, coral, sediment and glacial) are alternative, indirect climate observations which posses a recording characteristic \citep{Acevedo2015,Jones2004}. The best resolved features of climate recorded by proxies are of annual and seasonal cycles. However, the data recording process involved in such observations are very complex, encompassing physical, biological and chemical processes \citep{Evans2013}. Differentiating the climate and human impact on proxies is also a challenging work, for example, in terms of historical drought spells. The difficult challenge of inverting proxy records into climate information is traditionally done in the frame work of statistical modeling. However, multivariate linear regression techniques dominate this area \citep{Acevedo2015}. On the other hand, even using more sophisticated statistical models is problematic in a way that the overlapping time span between the instrumental records (weather station observations) and the proxy records are too 
short to train the statistical models. Climate model hindcast may serve as a modeling idea to study the long term climate variability. The climate models create dynamically consistent state of the climate using numerical methods. However, their reconstructed states are very sensible to the initial conditions and the forcing used in the model as well as parametrization schemes used for sub-grid scale processes (small processes which are not physically presented by model).

One of the novel and appealing approaches to reconstruct the past climate is Data Assimilation (DA) which blends the proxy records and the climate models \citep{Evensen2003,hughes2010,bronniman2011,Bhend20121,Hakim2013,hakim2014,Matsikaris2015,Hakim2016}.

\subsection{Challenges of Data Assimilation in climate studies}

For a review of DA techniques applied in climate studies (not the weather forecasting), I refer to the works of \cite{Acevedo2016,Steiger2015}. \cite{Acevedo2016,Steiger2015,hakim2014} showed that the climate models may not have forecasting skills longer than several years. This fact makes the usage of classical DA methods very limited. The models lose their skill shortly after initialization and will evolve freely till the next step where the observations (in form of proxies) are available. 

Considering the implementation ramification of DA methods in climate systems, their extremely computational expenses and the short forecasting skill of the models, an alternative DA approach (``off-line'' or ``no-cycling'') was applied by several scholars \citep{Acevedo2016,Steiger2015,Chen2015,hakim2014}. To my knowledge, there exists no study which applies such strategies in high resolution Regional Climate Model (RCM) simulations for long term climate studies. 

In this study, I will apply an Off-line DA (ODA) approach in a RCM simulation and test the performance of this method. I will present the problematic features of RCMs in simulating the climate and demonstrate the performance of very cheap statistical methods to correct the model's bias by means of ODA. 



\section{Data and Methods}
\label{sec:meth}
\subsection{Regional Climate Model}
The numerical RCM used in this study is the COSMO-CLM (CCLM) model version cosmo$\_$131108$\_$5.00$\_$clm8 \citep{Asharaf2012}. The horizontal resolution of the simulations is set to 0.44$^{\circ}$ $\times$ 0.44$^{\circ}$. The full model set-up file is included in the supplementary materials. ``The CCLM model is a non-hydrostatic RCM which uses terrain following height coordinates \citep{Rockel2008}, developed from the COSMO model, the current weather forecast model used by the German weather service (DWD)''\citep{Fallah2016}. In order to investigate the sensitivity of RCM to the initial conditions, two sets of simulations are conducted:\\

\begin{itemize}
\item \textbf{Default} (\textbf{NATURE}) simulation: 6 years long simulation over Europe driven by global atmospheric reanalysis data (6 hourly) produced by the European Centre for Medium-Range Weather Forecasts (ECMWF), the so-called ERAInterim \citep{Dee2011} (initial and boundary conditions are taken from ERAInterim). This run will be used as the ``Nature'' or ``True'' state of the climate in the investigations.
\item \textbf{Shifted} domain simulation: the same as \textbf{Default} but shifted 4 grid points to the Northwest compared with the \textbf{Default}.
\end{itemize}
\subsubsection{Model's Skill Metric}
\begin{figure}[!tbp]
\begin{center}

\includegraphics[width=1\linewidth]{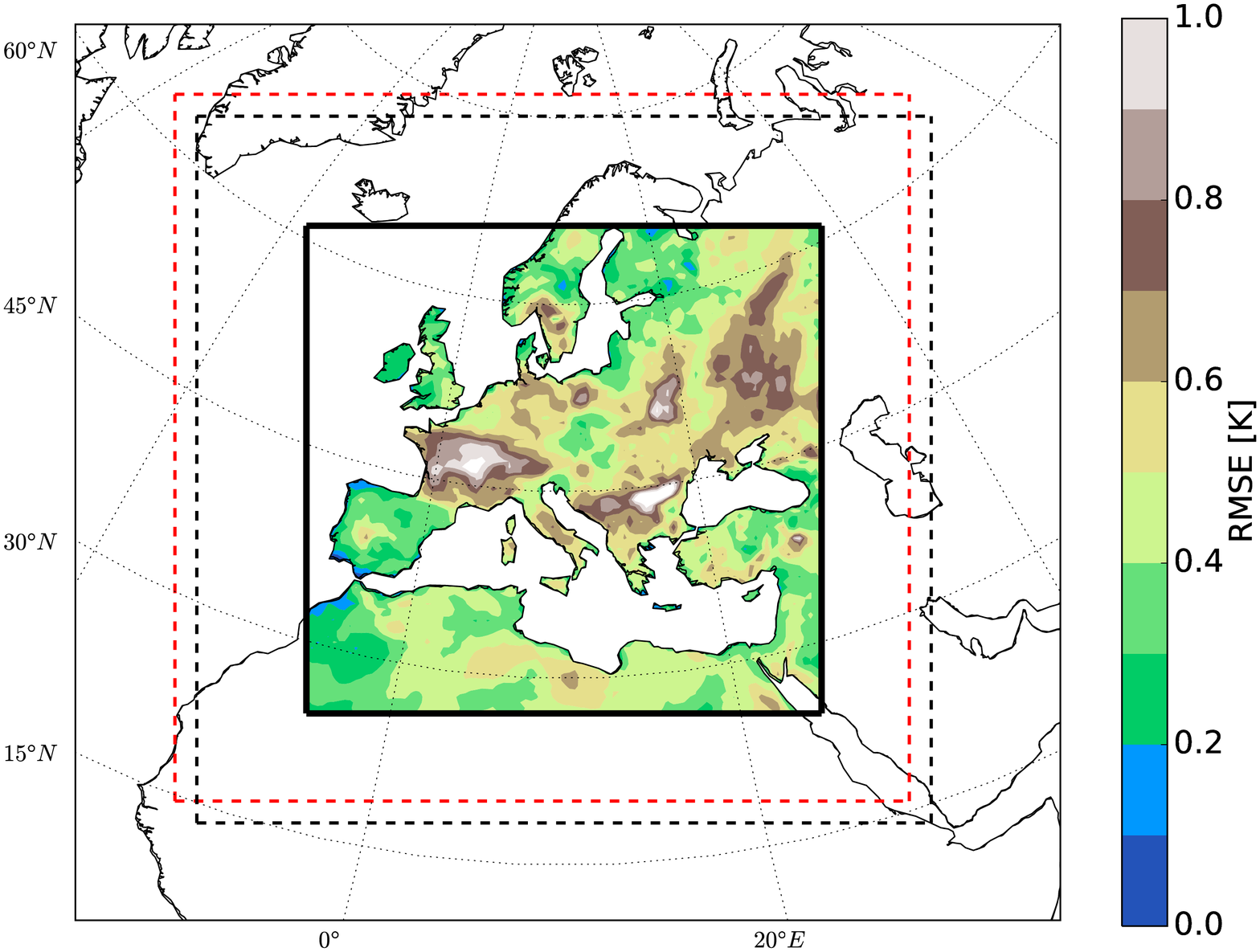}
\end{center}
\caption{RMSE (K) between the \textbf{NATURE} run and the \textbf{\emph{forecast}} of the \textbf{Shifted} run. Dashed black (red) box indicates the \textbf{NATURE} (\textbf{Shifted}) run. The bold box indicates the validation box. \label{fig:01}}
\end{figure}

The skill of the \textbf{\emph{forecast}} (or \emph{prior}) state of the \textbf{Shifted} simulation (without DA) can be measured, by the Root Mean Square Error (RMSE):

\begin{equation} \label{eq:rmse}
 RMSE(X^{Shifted}) = \left(\overline{\left(X^{NATURE} - X^{Shifted}\right)^2}\right)^\frac{1}{2} ,
\end{equation}
where $\overline{\phantom{gg}}$ denotes the time mean operator. Figure \ref{fig:01} shows the monthly near surface temperature (2 meter temperature) RMSE between the \textbf{Nature} and the \textbf{Shifted} simulations over the evaluation domain. For evaluation, the relaxation zone (the region where the boundary data is relaxed in the RCM, here 20 grid points next to the lateral boundaries) is removed. The last year of the simulations is used for validation and the first 5 years are considered as spin-up \footnote{In climate modeling Spin-up is the time considered for any model to reach a quasi equilibrium state.}. The 2 meter temperature (T2M) over oceans in CCLM model is interpolated from the driving Model (here ERAInterim) and not calculated by the model. Therefore the RMSEs over oceans are masked out from the analysis and only values over lands are shown.  

\subsection{Optimal Interpolation}

Prior to describing the DA methodology, I give a brief review of the Optimal Interpolation (OI) method  (for the full review see \cite{Gandin1966,Barth2008}). OI or ``objective analysis'' or ``kriging'' is one of the most commonly used and fairly simple DA methods applied since 1970s. The unknown state of the climate $X$ is the vector which has to be estimated based on the available observations ($Y$). Given the state vector $X$, the state on observations' location is obtained by an interpolation method (here nearest neighbor). This operation is noted as matrix $H$ and the state at the observations' location as $HX$. The ultimate goal is to find the nearest state to the ``True'' state of the climate ($X^{TRUE}$) the so-called \textit{Analysis} ($X^{A}$) given the observation ($Y$) and background  $X^{B}$ (first guess). The background and observation can be written as :

\begin{equation} \label{eq:1}
 X^{B} = X^{TRUE} + \eta^{B}
\end{equation}
\begin{equation} \label{eq:2}
 Y = HX^{TRUE} + \varepsilon
\end{equation}

where $\varepsilon$ and $\eta^{B}$ denote the observation and background errors, respectively.
\subsubsection{Assumptions in OI method}

In the applied OI scheme here, it is assumed that the background and observations are unbiased: 

\begin{equation} \label{eq:3}
E[\eta^{B}]=0
\end{equation}
\begin{equation} \label{eq:4}
E[\varepsilon]=0
\end{equation}

Other hypotheses are that the information about the observation and background errors are known (\textit{prior} knowledge) and they are independent:
\begin{equation} \label{eq:5}
E[\eta^{B}\eta^{B^{T}}]=P^{B}
\end{equation}
\begin{equation} \label{eq:6}
E[\varepsilon\varepsilon^{T}]=R
\end{equation}
\begin{equation} \label{eq:7}
E[\eta^{B}\varepsilon^{T}]=0
\end{equation}
\subsubsection{Analysis (\emph{posterior})}
The OI scheme is considered as the Best Linear Unbiased Estimator (BLUE) of the $X^{TRUE}$. BLUE has the following characteristics: 
\begin{enumerate}

 \item It is linear for $Y$ and $X^{B}$ 
 \item It is not biased: 
  \begin{equation} \label{eq:8}
    E[X^{Analysis}]=X^{NATURE}
  \end{equation}
 \item It has the lowest error variance (optimal error variance).
\end{enumerate}

The unbiased linear equation between $X^{B}$ and $Y$ can be written as : 

\begin{equation} \label{eq:9}
    X^{Analysis}=X^{B}+K(Y - HX^{B})
\end{equation}
where K is the ``Kalman gain'' matrix. Equation \ref{eq:9} can be written as :

\begin{equation} \label{eq:10}
    \eta^{Analysis}=\eta^{B}+K(\varepsilon - H\eta^{B}) = (I - KH)\eta^{B} + K\varepsilon
\end{equation}

Thus the error covariance of the \textbf{\emph{Analysis}} will be:
\begin{equation} \label{eq:11}
 P^{Analsis}(K)  = E[\eta^{Analysis}\eta^{Analysis^{T}}] = (I - KH)P^{B}(I - KH)^{T} + KRK^T
\end{equation}

The trace of matrix $P^{Analsis}$ indicates the error covariance of the analysis: 
\begin{equation}\label{eq:12}
\begin{aligned}
 Trace(P^{Analsis}(K))  = Trace(P^{B}) + Trace(KHP^{B}H^{T}K^{T}) \\
                        - 2Trace(P^{B}H^{T}K^{T})+ Trace(KRK^T)
\end{aligned}
\end{equation}

Given that the total error variance of \textbf{\emph{Analysis}} has its minimum value, a small $\delta K$ will not modify the total variance:
\begin{equation}\label{eq:13}
\begin{aligned}
 Trace(P^{Analsis}(K+\delta K)) - Trace(P^{Analysis}(K)) = 0 \\
  = 2Trace(KHP^{B}H^{T}\delta K^{T}) - 2Trace(P^{B}H^{T}\delta K^{T})+ 2Trace(KR\delta K^T) \\
  = 2Trace([K(HP^{B}H^{T} + R)]\delta K^{T})
\end{aligned}
\end{equation}

Given that the $\delta K$ is arbitrary, the Kalman gain is :
\begin{equation} \label{eq:14}
K = P^{B}H^{T}(HP^{B}H^{T}+R)^{-1}
\end{equation}
Finally, the error covariance of the BLUE is given by:
\begin{equation} \label{eq:15}
\begin{aligned}
P^{Analysis} = P^{B}-KHP^{B}\\
= P^{B}-P^{B}H^{T}(HP^{B}H^{T}+R)^{-1}HP^{B}
\end{aligned}
\end{equation}

In the scheme used here, $P^{B}$ is parametrized as following:
\begin{equation} \label{eq:16}
\begin{aligned}
P^{B}(x_{1},...,x_{n}, y_{1},...,y_{n}) = \sigma(x_{1},...,x_{n})^2 exp(-\frac{(x_1-y_1)^2}{{L^2}_{1}})... - \frac{(x_n-y_n)^2}{{L^2}_{n}}))
\end{aligned}
\end{equation}
Where the $\sigma(x_{1},...,x_{n})^2$ is the error variance and $L_n$ the correlation length (or scanning radius). The number of influential observations which contribute to each grid point has also to be given as the input variable.

\subsection{Observation System Simulation Experiment (OSSE)}
Models contain systematic errors which may have diverse origins (dynamical core, parametrization, initialization). DA schemes are also based on simplified hypotheses and are imperfect (e.g., here the Gaussian parametrization for $P^{B}$). All these sources of errors may also interact with one another in a way that tracing the source of problem may be impossible. In recent studies \citep{Acevedo2015, Acevedo2016} 
these error sources are neglected by using a simplified numerical experiment called OSSE.

\begin{figure}[!tbp]
\begin{center}
\includegraphics[width=1\linewidth]{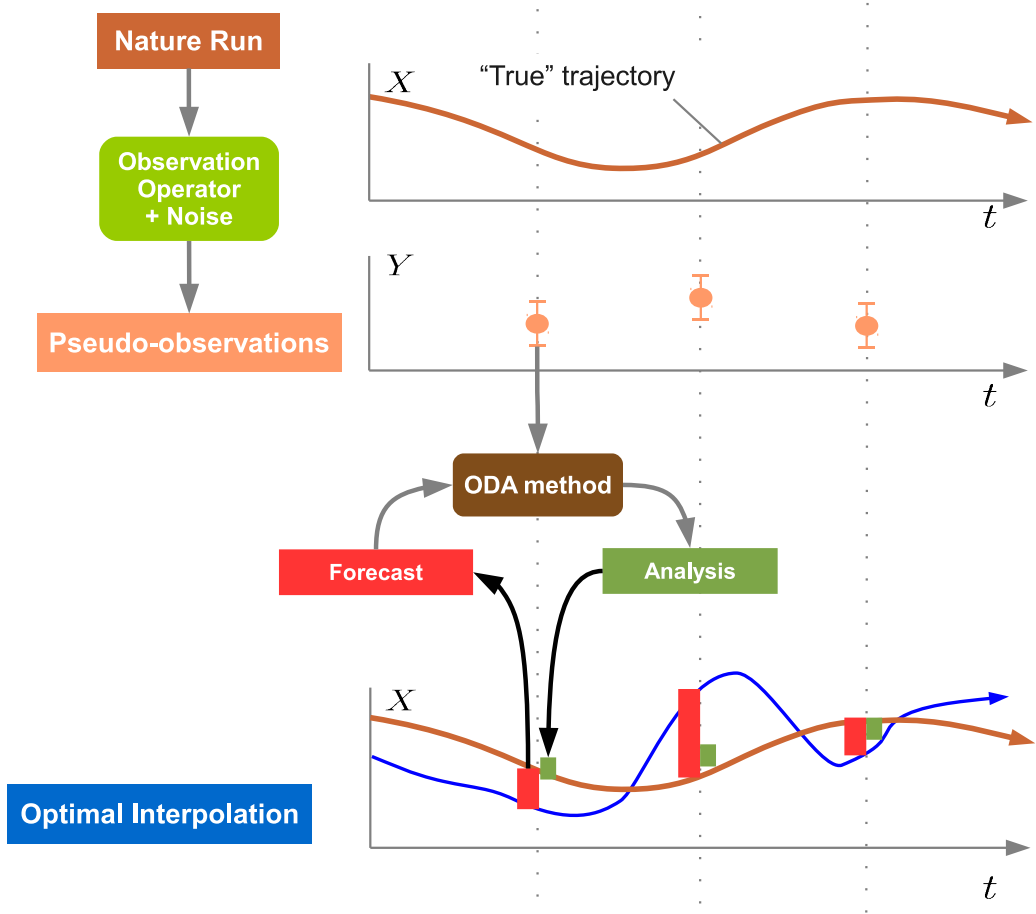}
\end{center}
\caption{Schematic of a typical Observation System Simulation Experiment (OSSE) with OI ODA method. \textit{t} designates the time axis and $X(Y)$ denotes the model state (observation) space. Sharp (rounded) cornered boxes represent data (processes). Red (green) vertical shadings indicate the \textit{Forecast} (\textit{Analysis}) RMSE. Vertical dotted lines represent the assimilation steps. Modified from \cite{Acevedo2016}. \label{fig:02}}
\end{figure}

The OSSE used in this study is described by the schematic shown in Figure \ref{fig:02}. First the \textbf{NATURE} model simulation $X^{NATURE}$ (``true'' run) is conducted as the final prediction target. Then using the output from \textbf{NATURE} run and adding random draws from a White Noise distribution (with $\mu = 0$ and $\sigma = 0.5$), the pseudo-observations are created which are interpolated over the observations' location. And finally, applying the OI scheme on the new \textbf{\textit{forecast}} (\textbf{Shifted}) run, the observationally constrained run $X^{DA}$ is obtained by assimilating the noisy observations.

\section{Results}
\label{sec:verify}
\begin{figure}[!tbp]
\begin{center}
\includegraphics[width=1\linewidth]{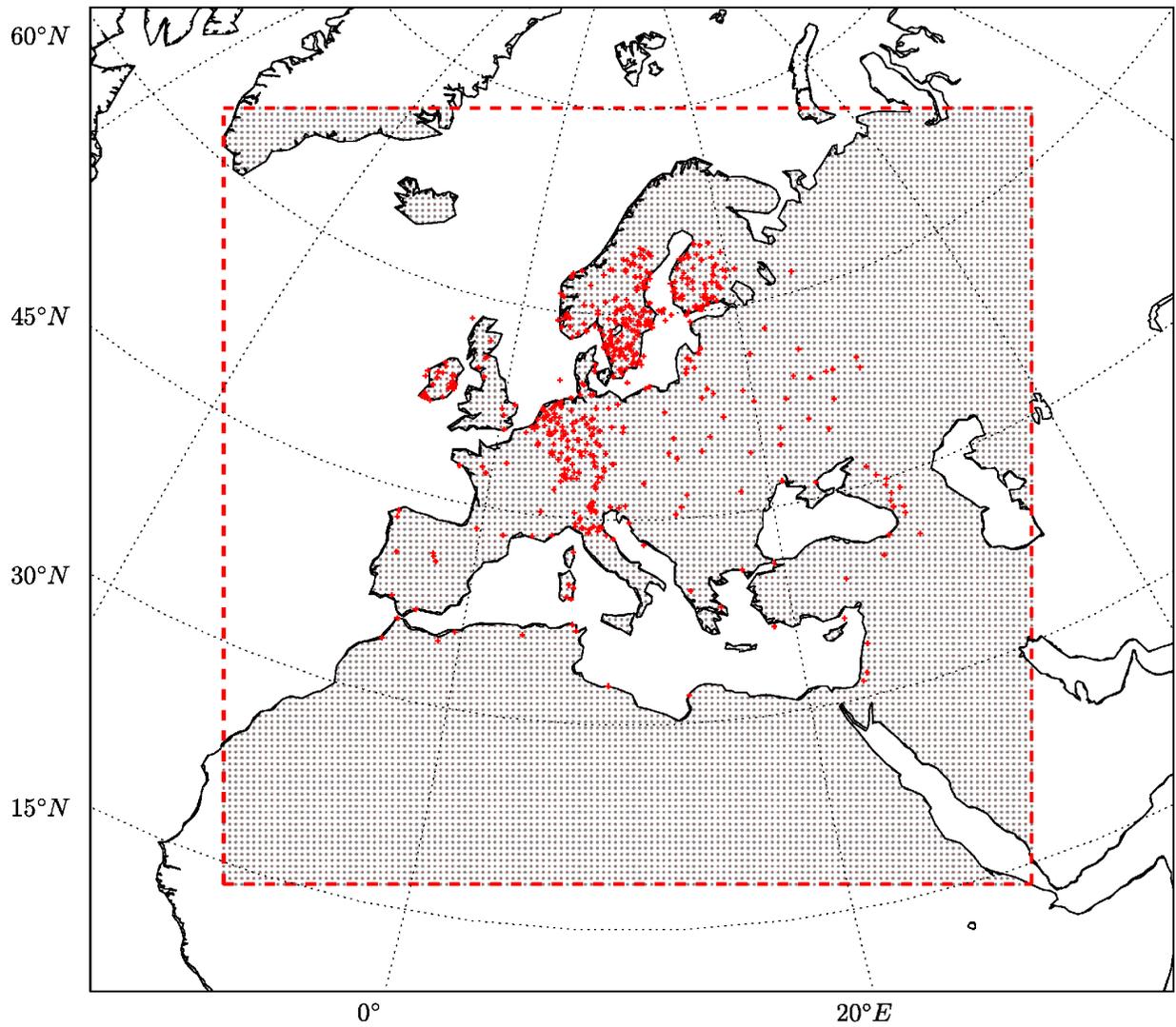}
\end{center}
\caption{Location of 500 random pseudo-observations (red crosses). Gray dots indicate the CCLM model's grid points over land. Red dotted box denotes the \textbf{Default} domain. \label{fig:03}}
\end{figure}

Comparing the two simulations exhibits that a slight change in the initial conditions of a RCM leads to large values (more than $1K$ in monthly values) of internal variability in the model \textbf{\textit{forecast}} (Fig.\ref{fig:01}). This feature has to be considered cautiously when conduction long-term climate simulations using RCMs, especially for the future climate projections. Here, I will show that the OI can significantly reduce the bias of the model \textbf{\textit{forecast}} by assimilating the pseudo-observations. The product of the OI is called \textbf{\textit{analysis}}, hereafter. Knowing the ``True'' state of the climate (\textbf{NATURE} run), the $\sigma(x_{1},...,x_{n})^2$ the error variance can be estimated. The location of 500 random meteorological stations of the ``ENSEMBLES daily gridded observational dataset for precipitation, temperature and sea level pressure in Europe called E-OBS'' \citep{Haylock2008} are used in this study to create the pseudo-observation data (Fig.\ref{fig:03}).

\begin{figure}[!tbp]
\begin{center}
\includegraphics[width=.7\linewidth]{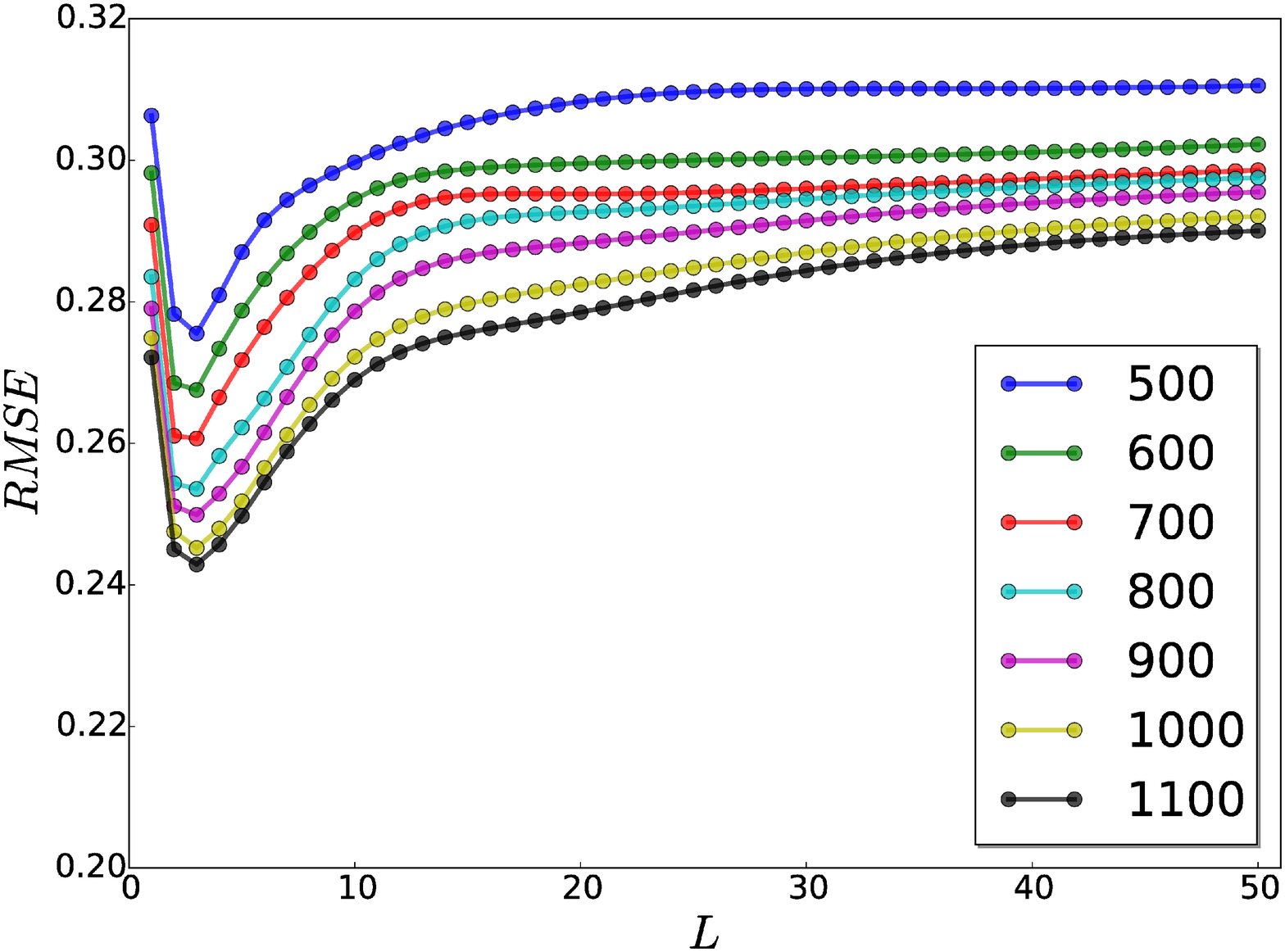}
\end{center}
\caption{Changes in averaged RMSE over the evaluation domain with respect to the changes of correlation length ($L$) and number of observations (different colors) . \label{fig:04}}
\end{figure}

In order to find the optimal correlation length ($L$), I have calculated the mean RMSE for the whole evaluation domain with several different correlation lengths ($1\times grid points\leq L \leq 50\times grid points$) as well as with different number of observations ($N=500,600,700,...,1100$). Figure \ref{fig:04} shows that the more the number of assimilated observations the lower the RMSE values are. The RMSE values show a minimum at correlation length of 3 (3$\times 0.44^{\circ}$ $\sim$ 150 $Km$) independent to the number of observations. Therefore, for this study the correlation length of three have been chosen. For long-term time smoothing (here I used monthly averaged temperatures), e.g., yearly or decadal time-averaging, the correlation length will increase \citep{Chen2015} accordingly.\par

\begin{figure}[!tbp]
\centering
\includegraphics[width=1\linewidth]{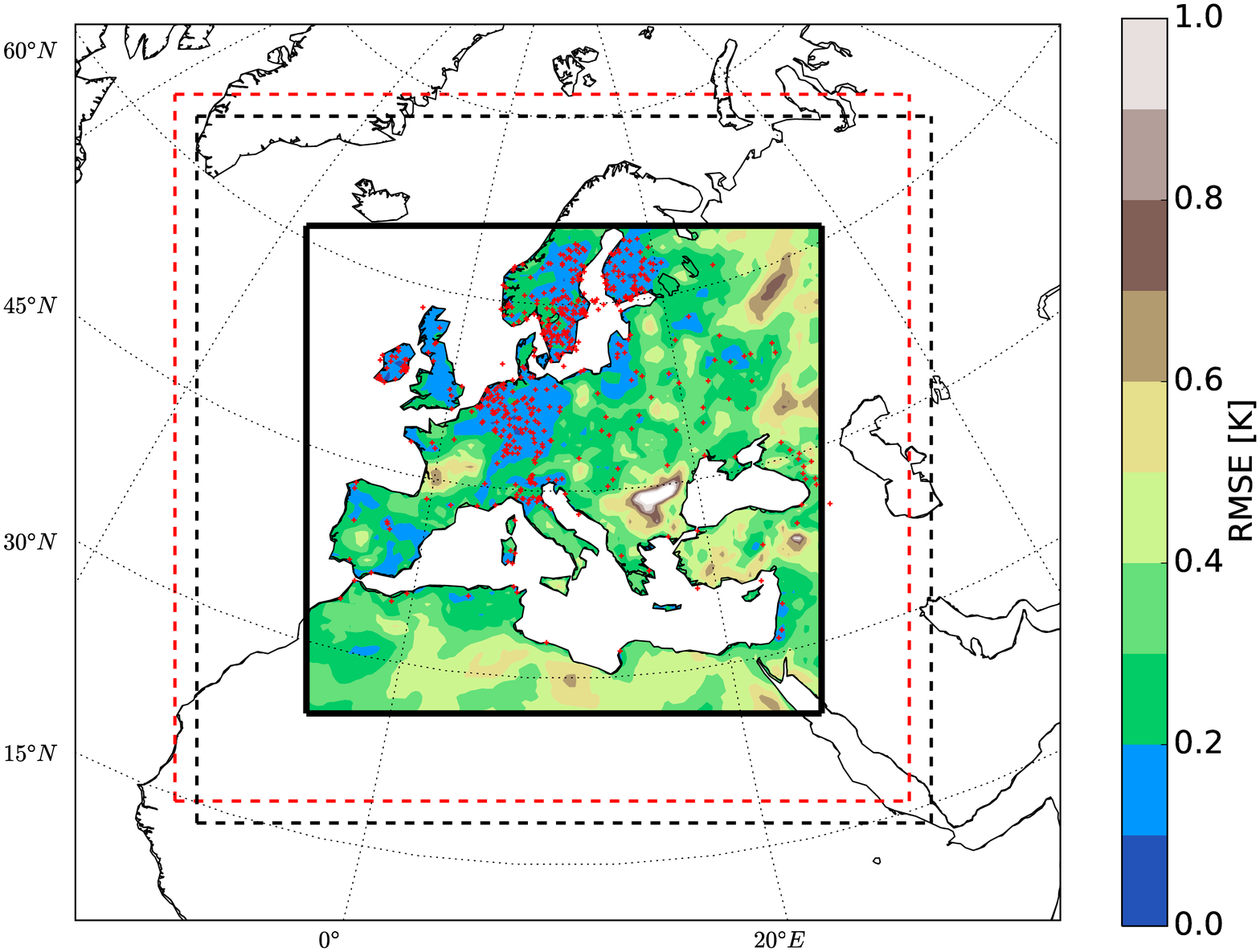}

\caption{RMSE of \textbf{\emph{Analysis}} with correlation length \textbf{3} and \textbf{500} assimilated observations. Locations of \textbf{500} random pseudo-observations are marked as red crosses. Dashed black (red) box indicates the \textbf{NATURE} (\textbf{Shifted}) run. The bold box indicates the validation box. \label{fig:05}}
\end{figure}

\begin{figure}[!tbp]
\centering
\includegraphics[width=1\linewidth]{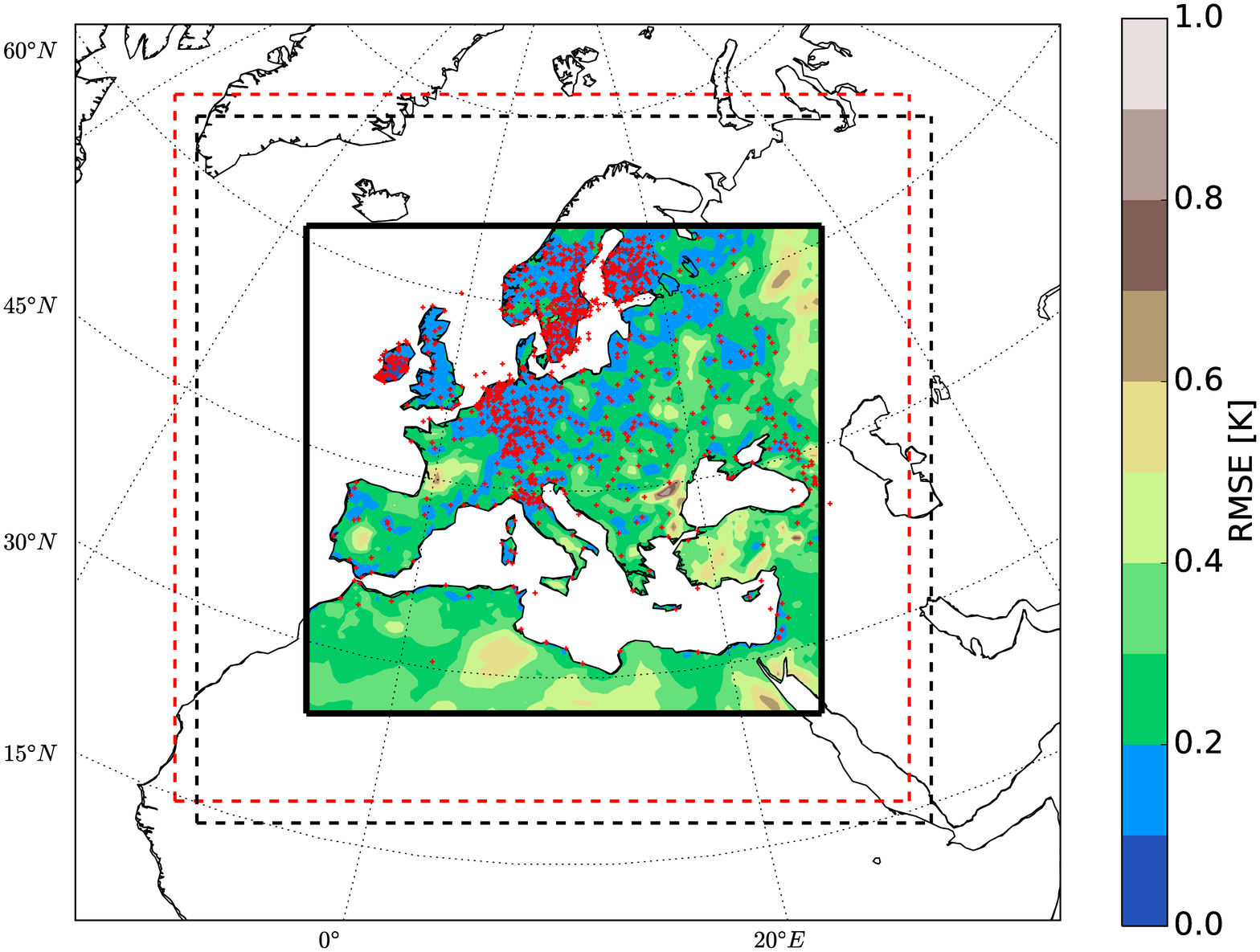}

\caption{RMSE of \textbf{\emph{Analysis}} with correlation length \textbf{3} and \textbf{1100} assimilated observations. Locations of \textbf{1100} random pseudo-observations are marked as red crosses. Dashed black (red) box indicates the \textbf{NATURE} (\textbf{Shifted}) run. The bold box indicates the validation box.\label{fig:06}}
\end{figure}

Figures \ref{fig:05} and \ref{fig:06} show the constrained simulations with 500 and 1100 pseudo-observations, respectively. The error reduction has maximum values around the observations' locations. Comparing Figures \ref{fig:01} and \ref{fig:06}, indicates that the OI method - despite its implementation easiness -  can significantly reduce the systematic error of the model.  

\section{Conclusion}
\label{sec:conc}
Using an OSEE, I conducted a simple numerical experiment to highlight the problematic features of RCMs in prediction of the near surface temperatures over Europe. By applying a fast and cheap DA method, I demonstrated that despite the easiness of OI, it can significantly reduce the bias of the regional model for monthly averaged values of near surface temperature. This method could potentially be applied for conducting long-term climate reanalysis data-sets with less complications than the classical DA methods. However, for a real application of this method, there exist several open questions: (i) is there enough proxy data available for the study region (for example less coverage over Central Asia)? (ii) how is the accuracy of the data in value and timing? (iii) with which data we shall drive the RCM (for initialization and boundary conditions)?\par

The simulations in this study are driven with ERAInterim which is produced by sophisticated methodologies combining available observations and models for the recent time (last 40 years). For the past climate simulations which go beyond the time-span of observations (i.e., for time spans greater than several centuries), the uncertainties in the model initialization and forcing (i.e., radiation, anthropogenic, volcanic) will increase significantly. Therefore, for such studies usage of ensemble simulation approach (different models, different initializations, different forcing) is of important interest and should be considered. However, by applying an ensemble of simulations the complexity and implementation of data analysis will increase exponentially. Such a cheap and fast method presented here, may pave the wave for handling these big data-sets in a way that the models' bias are corrected by available observations. 

\bigskip
\begin{center}
{\large\bf SUPPLEMENTARY MATERIAL}
\end{center}

\begin{description}

\item[Title:] Open Access codes for analysis and plotting

\item[Open Source Codes:] All the applied codes in this research and the CCLM model set-up files are uploaded and hosted at \url{https://github.com/bijanfallah/historical_runs} along with a README file. The scientific codes are written in Python, BASH and GNU Octave under GNU GENERAL PUBLIC LICENSE. For any question or feedback please do not hesitate to contact me by email: info@bijan-fallah.com .

\item[Optimal Interpolation Code:] The optimal interpolation Fortran module with Octave interface of GeoHydrodynamics and Environment Research (GHER) is used. The code and the documentation can be obtained at their website:\\ (\url{http://modb.oce.ulg.ac.be/mediawiki/index.php/Optimal_interpolation_Fortran_module_with_Octave_interface}).

\end{description}



\bibliographystyle{agsm}

\end{document}